\documentstyle[prd,aps]{revtex}
\begin{document}
\textheight=640pt
\textwidth=440pt
\topmargin=-25pt
\leftmargin=-25pt
\hoffset -0.4cm

\title{Quark rotation asymmetry and baryon magnetic moments}
\author{W. Li\thanks{E-mail: liw@iopp.ccnu.edu.cn} 
    and X. Cai\thanks{E-mail: xcai@wuhan.cngb.com} \\ 
   \footnotesize \sl Institute of Particle Physics, Hua-zhong Normal
                     University, Wuhan, 430079, P. R. China}

\maketitle
\vskip 0.5cm
\begin{abstract}
Quark rotation asymmetry is proposed in calculating baryon magnetic moments
. After taking into account interactions enforced on constituent
quarks, assumed to be linear and Coulomb potentials, respectively,
and the quark rotation asymmetry, we fit the theoretical values, based on two more 
hypotheses and several other reasonable assumptions, of baryon magnetic moments
with those from experiments. The good fitting results shows the necessity of
the consideration of quark rotation asymmetry within baryon.

\vspace{0.5cm}
\mbox{} \hspace{-0.565cm} PACS number(s): 24.85.$+$p, 14.20.$-$c, 13.40.e
\end{abstract}

\twocolumn
\hoffset 0.5cm
\rightskip -0.2cm
\begin{center}
\section*{\bf I. Introduction}
\end{center}

Problem on the spin of proton has been discussed since a long time
ago. Physicists are still still puzzled with where proton gets its spin
despite that efforts have been made.

As known, in Naive Quark Model (NQM), proton's spin is assumed to be 
carried by three valence quarks. So the magnetic moments of 
proton is entirely contributed also by three valence quarks. The baryon
magnetic moments presented by NQM fit the experimental measurement of baryon
magnetic moments well. So for a quite long time our knowledge of 
proton spin stayed comfortably on the level of NQM.

However, in the last few years, the well-known ``Spin-Crisis'', which is
arisen by the polarized experiments of muon scattering from proton at CERN
by the European Muon collaboration (EMC) [1],
has led to the new argument of the origin of proton spin. These experiments
presented the surprising conclusion that only a small part of proton spin
is carried by the spin of light quarks (and antiquarks) it contains, which
was in disagreement with the popular picture for nucleon structure.

Several attempts [2,3] have been made in trying to solve the puzzle
mentioned above. Some authors [2] argued that
the orbital angular momentum of constituent quarks may have significant
contribution to proton spin. But other authors [3] argue that the data obtained
by EMC give no definite conclusion on proton spin content and
errors, considering the fact that effect resulting from the uncertainty of
extrapolation had not been fully considered. The argument is still on. Thus
the EMC data and the works [1-3] excite the old question ``May the spin
of baryon be attributed to the rotation of constituent quarks within 
baryon ?''. It is worth discussing.

In our opinion it is quite natural to accept the concept of ``rotating proton''.
Actually that concept has been emphasised by Chou and Yang [4] in 1974.  In
1974, Sehgal [5] discussed the importance of constituent quarks rotation.
And Ellis-Jaffe sum rule [6] was also given in the same year. Since
then many papers, e.g., Refs. [7-9], discussed the origin of proton spin. 
In Ref. [9] Meng
proposed two polarization experiments to test the possible existence for
rotating constituents. In a recent work [10] presented by Casu and Sehgal
a model with collective quark rotation is used in discussing proton spin
and baryon magnetic moments. Their results fit the experimental ones well.
Therefore, the above works show the possible contribution from quark
revolving to baryon spin and magnetic moments.

In this work we adopt the concept of ``revolving quark''. We consider the possible
existence of quark rotation asymmetry. We simply assume the interaction
potentials enforced on constituents to be linear
and Coulomb type respectively. With the formulae for baryon
magnetic moments, we fit baryon magnetic moments with those from
experiments.

In section II the formulae for contribution of quark moments to baryon
magnetic moments are derived quantum mechanically, 
in section III the contribution from orbital
angular momentum to baryon magnetic moments through a simple calculation 
, and the fitted results, are presented. The last section is a brief 
conclusion.  

\vskip 1cm
\begin{center}
\section*{\bf II. FORMULAE FOR CONTRIBUTION FROM QUARK MOMENTS TO
	 BARYON MAGNETIC MOMENTS}
\end{center}

In this section we will show formulae in calculating baryon magnetic moments
without considering the contribution from constituent quarks rotation.

First of all, the $z$ component of polarized quark and antiquark contribution to
the spin of a polarized proton is [5] $\langle S_{z} \rangle =(\Delta u+\Delta d
+ \Delta s) /2 $, with $\Delta q$ the net polarization of quarks of flavor $q$, and
\begin{eqnarray} \nonumber
\Delta q &=&\int dx [q_{+}(x)-q_{-}(x)]  \\
&+& \int dx[\bar{q}_{+}(x)-\bar{q}_{-}(x)]  
\end{eqnarray}

\noindent
with $q_{\pm}(\bar{q}_{\pm})$ being the densities of parton quarks
(antiquarks) with helicities $\pm \frac {1}{2}$ in a proton with
helicity $\frac {1} {2}$. $\Delta u$, $\Delta d$ and $\Delta s$ are known as parton spins, which can
be related to the values of the axial vector coupling constants $G_{A}$,
$a^{(8)}$ and $\langle S_{z} \rangle$ through the following formulae
\begin{eqnarray} \nonumber
\Delta u &=&\frac {2}{3} \langle S_{z} \rangle+\frac {1}{2} G_{A}+\frac {1}{6}
a^{(8)}  ,\\  \nonumber
\Delta d &=&\frac {2}{3} \langle S_{z} \rangle-\frac {1}{2} G_{A}+\frac {1}{6}
a^{(8)}  ,\\
\Delta s &=&\frac {2}{3} \langle S_{z} \rangle-\frac {1}{3} a^{(8)}  .
\end{eqnarray}

\noindent
We take $G_{A}$ and $a^{(8)}$ their experimentally measured values,1.26 and 0.60,
respectively, and $\langle S_{z} \rangle$ is used as a fitting parameter.

To a first approximation, the baryon magnetic moments can be expressed as a
vector sum of the quark moments plus a contribution from any orbital angular
moment of the quarks [11]. In this section we will only present the formulae
of the contribution from quark moments to baryon magnetic moments. The
orbital angular moment will be added in the next section.

The magnetic moments operator $\hat {\mu}_B$ of a baryon is given by 
\begin{equation}
\hat {\mu}_B=\sum\limits_{i=1}^{3} \mu_q(i) \hat {\sigma}(i)
\end{equation}
\noindent
, where the sum is over the three quarks in a baryon, 
and $\hat {\sigma}$ is the Pauli matrix. The quark moments of any baryon is
the expection value of $\mu_3$ (the third or Z component of $\vec
{\mu_B}$ with respect to a baryon wave function $\Psi_B$ which is maximally
polarized along the Z-axis, that is [11], 
\begin{equation}
\mu_B=\langle \Psi_B \vert \sum\limits_{i=1}^{3} \mu_q(i) \hat {\sigma}(i)
\vert \Psi_B \rangle
\end{equation}

It is hence possible to evaluate $\mu_B$ in terms of $\mu_q$ for any baryon
once the flavor and spin wave functions of the baryon are specified. Using
Eq. (4) , after some careful calculation, one can write down the contribution
of quark momnets to baryon magnetic moments for spin-$\frac {1} {2}$ baryon
Octet as follows [12,13], 
\begin{eqnarray}    \nonumber
\underline {\mu}(p) &=&\mu_{u} \delta u +\mu_{d} \delta d +\mu_{s} \delta s  ,\\ \nonumber
\underline {\mu}(n) &=&\mu_{u} \delta d +\mu_{d} \delta u +\mu_{s} \delta s  ,\\  \nonumber
\underline {\mu}(\Sigma^{+}) &=&\mu_{u} \delta u +\mu_{d} \delta s +\mu_{s} \delta
d  ,\\ \nonumber
\underline {\mu}(\Sigma^{-}) &=&\mu_{u} \delta s +\mu_{d} \delta u +\mu_{s} \delta
d  ,\\ \nonumber
\underline {\mu}(\Xi^{-}) &=&\mu_{u} \delta s +\mu_{d} \delta d +\mu_{s} \delta
u  ,\\ \nonumber
\underline {\mu}(\Xi^{0}) &=&\mu_{u} \delta d +\mu_{d} \delta s +\mu_{s} \delta
u  ,\\ \nonumber
\underline {\mu}(\Lambda^{0}) &=&\frac {1}{6}(\delta u+4\delta d +\delta s)
(\mu_{u}  \\ \nonumber
&&+\mu_{d})+\frac {1}{6} (4\delta u -2\delta d \\  \nonumber
&&+4\delta s) \mu_{s}  ,\\  \nonumber
\underline {\mu}(\Sigma^{0}) &=&- \frac {1}{2 \sqrt{3}} (\delta u -2\delta d
+ \delta s)\\
&&\times (\mu_{u}-\mu_{d})  .
\end{eqnarray}
\noindent
, where, $\delta q$ is defined as
\begin{eqnarray}  \nonumber
\delta q&=&\int dx [q_{+}(x)-q_{-}(x)]  \\
&-&\int dx[\bar{q}_{+}(x)-\bar{q}_{-}(x)]  ,
\end{eqnarray}
\noindent
which differs from expression of $\Delta q$ only in the sign of the 
antiquark contribution.

Relationship between $\Delta q$ and $\delta q$ is set through two reasonable
hypotheses. These two hypotheses are based on the consideration of nucleon 
structure, hence they may be reasonable and acceptable.
 
\vspace{0.4cm}
\noindent
{\bf Hypothesis} \hspace{0.3cm}I

\vspace{0.3cm}
One may expect that sea quarks in a polarized baryon entirely result
from a cloud of spin-zero mesons. Such models have been discussed, for
instance, in Ref. [14]. In such case $\Delta q=\delta q$.

\vspace{0.4cm}
\noindent
{\bf Hypothesis} \hspace{0.3cm}II

\vspace{0.3cm}
One may also expect that sea quarks in a polarized baryon are entirely produced
by gluons splitting $g\rightarrow q\bar{q}$. For simplicity, one may
let $\bar{u}_{+}-\bar{u}_{-} \simeq \bar{d}_{+}-\bar{d}_{-}
\simeq k(\bar{s}_{+}-\bar{s}_{-}) \simeq k(s_{+}-s_{-})$, where
$k$ represents the relative abundance of various antiquarks within the
baryon. For $k=1$ it is the case in Ref. [12]. Generally $\delta u
=\Delta u-k\Delta s, \delta d=\Delta d-k\Delta s, \delta s=0$. In our work
$k=0.5$ is used.

Consider all the above things and the additional two relationship, 
that is, $\mu_{u}=-2\mu_{d}$ 
and $\mu_{s}= 3\mu_{d}/5$, one can reexpress contribution of quark moments
to baryon megnetic moments in terms of parameters $\mu_{u}$, $G_{A}$, 
$a^{(8)}$ and $\langle S_{z} \rangle$, during which only $\mu_{u}$ and 
$\langle S_{z} \rangle$ are undetermined. 
 
\vskip 1cm

\begin{center}
\section*{\bf III. THE POSSIBLE EXISTENCE OF QUARK ROTATION ASYMMETRY}
\end{center}

Formulae in last section for quark moments are given without
considering the contribution from the revolving of constituent quarks.
In this section we will add the contribution from quark revolving to baryon
magnetic moments, considering the possibility of quark rotation asymmetry.
Before doing this let us briefly review some other works.

In Ref. [10] the authors proposed a collective rotation method in calculating
the spin of proton and baryon magnetic moments. The model they used is a
flux-string one. In their model the quarks will tend to be situated at the
corners of an equilateral triangle in the plane transverse to proton spin.
They assume that the correlated 3-quark structure rotates collectively
around the $z$-axis, which adds contribution of the revolving quarks
to proton spin and to baryon magnetic moments. Some reasonable results are
obtained. 

As we know the total angular momentum of a polarized proton can be resolved
as $J_{z}=\langle S_{z} \rangle +\langle L_{z} \rangle +\Delta G=1/2$.
$\Delta G$ is the contribution from gluons. Here $\langle L_{z} \rangle$
is related to the motion of quarks within baryon and shared by all 
the constituents. In Ref. [10] the authors let the revolving radius r
be the same for each quark, so the orbital angular momentum each quark 
contributes is proportional to its mass.

We appreciate the simplicity of the ideas presented in Ref. [10]. However,
we would like to emphasize here we should pay more attention to the
structure of baryons. It is most important. 

We should note there are three valence quarks within baryons,
 which mainly contribute to orbital angular momentum. If baryons are made
up of three same quarks there will exist symmetry in quark rotation 
without any doubt. However the three quarks within baryons are not
exactly same. In proton there are two $u$ quarks and one $d$ quark, for
neutron there are two $d$ quarks and one $u$ quark. Even for $\Lambda^{0}$
there exists one $u$, $d$, and $s$ quark respectively. At least two quarks
within baryons are different. We know that $u$, $d$, and $s$ quark
differ from each other in many aspects, during which most important one
is that they have different mass. The strange quark is much more heavier
than the up and down quarks. Due to the difference, the rotation symmetry is
badly breaking. So we propose the possible effect of constituent
quark rotation asymmetry. But the concept ``asymmetry'' is what we should
fully understand. The constituents still rotate along the geometry center
of the triangle composed by three quarks. But due to the mass difference
between constituents the triangle is scalene instead of equilateral, which
means the length of radius $r$ for different quark when they rotate along 
the axis is different. But another question arises ``How can the radius be
determined?''.

To answer this question, one has to add some more assumptions.
It is obvious that the assumed centripetal force for quark revolving
is entirely provided by the interaction between constituents. Since
no one can find out the exact form of the interaction potential between
quarks inside a hadron, one can and should assume the potential
to be in the simplest form. In this paper, 
the total interactions enforced on a quark from other quarks within baryon
is assumed to be proportional or inversely proportional to the 
revolving radius of the quark. The first interaction form corresponds to
a uniform color tube, and the second one is a Coulomb-like one. 
Sure, they are simplest considerations about internal quark interactions.
It can also be assumed that the direction of the force acting on a revolving
quark points to the center. With these assumptions one can easily calculate
the revolving quark contribution to baryon magnetic moments. 
 
In the first case we consider a quark in a linear potential, say $U=cr$, with $c$ 
a positive number and $r$ the radius for quark of mass $m$
revolving around the axis. So the force acting on the quark is a constant, $F=-c$. 
Due to the flavor independence of color interactions, the forces 
acting on all revolving quarks are the same for every valence quark within
the baryon. Then using $F=m \omega^{2} r$ one can get the dependence of
$r$ on $m$: $r \propto 1/m$. Hence the orbital angular momentum contributed from
quark $q_{i}$ of mass $m_{i}$ is $[(1/m_{i})/(1/m_{1}+1/m_{2}+1/m_{3})]
\langle L_{z} \rangle$. Adding the revolving quark contribution to baryon
magnetic moments in Eq. () one can obtain the entire formulae of baryon
magnetic moments when the interaction potential between constituents is
assumed to be a linear one,
\begin{eqnarray}  \nonumber
\mu_{L}(p) &=&\underline {\mu}(p) +[2\mu_{u} \times \frac {1}{3}\\   \nonumber
&&+\mu_{d} \times \frac {1}{3}] \langle L_{z} \rangle  ,\\ \nonumber
\mu_{L}(n) &=&\underline {\mu}(n) +[2\mu_{d} \times \frac {1}{3} \\  \nonumber
&&+\mu_{u} \times \frac {1}{3}] \langle L_{z} \rangle  ,\\  \nonumber
\mu_{L}(\Sigma^{+}) &=&\underline {\mu}(\Sigma^{+})+[2\mu_{u} \times \frac {1}{\lambda +2}\\  \nonumber
&&+\mu_{s} \times \frac {\lambda}{\lambda +2}] \langle L_{z} \rangle  ,\\ \nonumber
\mu_{L}(\Sigma^{-}) &=&\underline {\mu}(\Sigma^{-})+[2\mu_{d} \times \frac {1}{\lambda +2} \\  \nonumber
&&+\mu_{s} \times \frac {\lambda}{\lambda +2}] \langle L_{z} \rangle  ,\\ \nonumber
\mu_{L}(\Xi^{-}) &=&\underline {\mu}(\Xi^{-}) +
[2\mu_{s} \times \frac {\lambda}{2\lambda +1} \\ \nonumber
&&+\mu_{d} \times \frac {1}{2\lambda +1}] \langle L_{z} \rangle  ,\\ \nonumber
\mu_{L}(\Xi^{0}) &=&\underline {\mu}(\Xi^{0})
+[2\mu_{s} \times \frac {\lambda}{2\lambda +1}  \\  \nonumber
&&+\mu_{u} \times \frac {1}{2\lambda +1}] \langle L_{z} \rangle  ,\\ \nonumber
\mu_{L}(\Lambda^{0}) &=&\underline {\mu}(\Lambda^{0}) +
[\mu_{u} \times \frac {1}{\lambda +2}\\    \nonumber
&&+\mu_{d} \times \frac {1}{\lambda +2} \\  \nonumber
&&+\mu_{s} \times \frac {\lambda} {\lambda +2} ] \langle L_{z} \rangle  ,\\ \nonumber
\mu_{L}(\Sigma^{0}) &=&\underline {\mu}(\Sigma^{0})
+[\mu_{u} \times \frac {1}{\lambda +2}\\   \nonumber
&&+\mu_{d} \times \frac {1}{\lambda +2} \\
&&+\mu_{s} \times \frac {\lambda} {\lambda+2}] \langle L_{z} \rangle  .
\end{eqnarray}

\noindent
where $\lambda$=$m_{d}/m_{s}$=$0.6$ and the subscript $L$ indicates linear potential.
We then can fit the corrected baryon magnetic moments with experimental values
of baryon. In this paper, we fit the baryon magnetic moments
with two different hypotheses presented earlier about sea quark contribution
respectively. To every hypothesis we perform two fits. 
In fit 1, we let $\mu_{u}$,
$\langle S_{z} \rangle$ as fitting parameters with the constraint $\langle S_{z}
\rangle +\langle L_{z}\rangle =1/2$. This is, in fact, the extreme hypothesis that
the ``missing'' angular momentum of the proton is precisely accounted for
by the orbital angular momentum of revolving quarks. In fit 2, 
 we let $\mu_{u}$, $\langle S_{z} \rangle$ and$\langle L_{z}
\rangle$ as fitting paprmeters, where $\langle S_{z} \rangle$ and $\langle L_{z}
\rangle$ are free. The fitting results are given in Table I.

For the second assumed potential, the interaction between constituent quarks
is of Coulomb type, $U=c/r$, and one can get the dependence of
$r$ on $m$: $r \propto 1/ \sqrt [3]{m}$. Hence the orbital angular momentum
carried by the quark $q_{i}$ of mass $m_{i}$ is $[\sqrt [3]{m_{i}}/(\sqrt
[3]{m_{1}}+\sqrt [3] {m_{2}} +\sqrt [3]{m_{3}})] \langle L_{z} \rangle$.
With the revolving quark corrections taken into consideration in Eq. (2), one
can get the corrected formulae of baryon magnetic moments
\begin{eqnarray}      \nonumber
\mu_{C}(p) &=&\underline {\mu}(p) +[2\mu_{u} \times \frac {1}{3}\\ \nonumber
&&+\mu_{d} \times \frac {1}{3}] \langle L_{z} \rangle  ,\\ \nonumber
\mu_{C}(n) &=&\underline {\mu}(n) +[2\mu_{d} \times \frac {1}{3} \\ \nonumber
&&+\mu_{u} \times \frac {1}{3}] \langle L_{z} \rangle  ,\\ \nonumber
\mu_{C}(\Sigma^{+}) &=&\underline {\mu}(\Sigma^{+}) +
[2\mu_{u} \times \frac {\lambda^{*}}{2\lambda^{*} +1}\\ \nonumber
&& +\mu_{s} \times \frac {1}{2\lambda^{*} +1}] \langle L_{z} \rangle  ,\\ \nonumber
\mu_{C}(\Sigma^{-}) &=&\underline {\mu}(\Sigma^{-})
+[2\mu_{d} \times \frac {\lambda^{*}}{2\lambda^{*} +1} \\ \nonumber
&&+\mu_{s} \times \frac {1}{2\lambda^{*} +1}] \langle L_{z} \rangle  ,\\  \nonumber
\mu_{C}(\Xi^{-}) &=&\underline {\mu}(\Xi^{-}) +
[2\mu_{s} \times \frac {1}{\lambda^{*} +2} \\  \nonumber
&&+\mu_{d} \times \frac {\lambda^{*}}{\lambda^{*} +2}] \langle L_{z} \rangle  ,\\ \nonumber
\mu_{C}(\Xi^{0}) &=&\underline {\mu}(\Xi^{0})
+[2\mu_{s} \times \frac {1}{\lambda^{*} +2} \\  \nonumber
&&+\mu_{u} \times \frac {\lambda^{*}}{\lambda^{*} +2}] \langle L_{z} \rangle \\  \nonumber
\mu_{C}(\Lambda^{0}) &=&\underline {\mu}(\Lambda^{0}) +
[\mu_{u} \times \frac {\lambda^{*}}{2\lambda^{*} +1} \\ \nonumber
&&+\mu_{d} \times \frac {\lambda^{*}}{2\lambda^{*} +1}  \\ \nonumber
&&+\mu_{s} \times \frac {1} {2\lambda^{*} +1} ] \langle L_{z} \rangle  ,\\  \nonumber
\mu_{C}(\Sigma^{0}) &=&\underline {\mu}(\Sigma^{0})
+[\mu_{u} \times \frac {\lambda^{*}}{2\lambda^{*} +1} \\ \nonumber
&&+\mu_{d} \times \frac {\lambda^{*}}{2\lambda^{*} +1} \\
&&+\mu_{s} \times \frac {1} {2\lambda^{*}+1}] \langle L_{z} \rangle  ,
\end{eqnarray}

\noindent
where $\lambda^{*}$= $\sqrt [3]{m_{d}/m_{s}}$=$\sqrt [3]{0.6}$ and the
subscript $C$
indicates Coulomb potential. Similarly we fit the corrected baryon
magnetic moments with experimental values of baryon with hypotheses
assumed before respectively. The fitting procedures and constraints are exactly
the same as for the case with linear potential.
Fitted results are presented in Table II.

\vskip 1cm
\begin{center}
\section*{\bf IV. CONCLUSIONS}
\end{center}

In this work we discuss the possible quark rotation asymmetry within baryon.
Based on two kinds of hypotheses on sea quark contribution and
some reasonable assumptions about the internal interactions, the
baryon magnetic moments are calculated. It should be pointed out that
contribution from quark moments to baryon magnetic moments can be 
calculated quantum mechanically,
while the contribution from angular momentum is a quasiclassical
approximation. We fit the baryon magnetic moments with the experimental
ones. The fitted results seem consistent
with experimental measurements to certain extent. Thus we may possibly provide
a simple method in solving the question ''Where does the proton get its spin?
'' [15].

In our work we assume interactions between constituents are
linear or Coulomb potential. To some extent the assumptions are too
simple. We know that quarks are confined in the baryon.
From the study of hadron spectrum it is known that the potential between
quarks can be approximately written as $Ar+B/r$. Such potential can provide
some good description of baryon feature. One can see that our two choices are
just the two limits of $Ar+B/r$ when parameter $B$ and $A$ are assumed
very small, respectively. So the two forms of interactions can be thought
as reasonable.

In general, a clear picture of revolving quark within baryon is presented
in this paper. Calculation in this work is not too difficult.

One should also note that from our work and others parton
quark model can give a good description of baryon magnetic moments, 
axis vector coupling and the proton spin. It is sure that we need more data
to test the feasibility of quark parton model. 

By now the inside structure of baryon is still unknown to us. We expect
that future experiments on LHC can provide some picture in the nucleon
structure, which would be of interest to both 
theoretical and experimental physicists.

This work was supported by the National Science
Foundation of China, the Natural Science Foundation of Hubei,
the Foundation of the State Education Commission of China.

\onecolumn
\section*{\normalsize \bf  TABLE I. Fits of magnetic moments when quark interactions are linear
potential.}

\begin{tabular}{c|c|c|c|c|c} \hline
&Experimental& \multicolumn{2}{|c|}{Hypothesis I}& \multicolumn{2}{|c}{Hypothesis II}\\  \cline{3-6}
&Data&fit 1 &fit 2 &fit 1 & fit 2 \\ \hline
$\mu_L (p)$&$ 2.79 \pm 0.10$ &2.75 & 2.66 &2.75 & 2.79 \\
$\mu_L (n)$&$ -1.91 \pm 0.10$ &$-1.82$ & $-1.92$ &$-1.85$ & $-1.74$\\
$\mu_L (\Sigma^{+})$&$ 2.46 \pm 0.10 $&2.71 & 2.61 &2.63 & 2.58\\
$\mu_L (\Sigma^{-})$&$ -1.16 \pm 0.10$ &$-1.09$ & $-1.17$ &$-1.03$ & $-1.12$\\
$\mu_L (\Xi^{-})$&$ -0.65 \pm 0.10$&$-0.48$ & $-0.56$& $-0.42$ & $-0.51$\\
$\mu_L (\Xi^{0})$&$-1.25 \pm 0.10$&$-1.24$ & $-1.34$& $-1.34$ & $-1.37$\\
$\mu_L (\Lambda^{0})$&$-0.61 \pm 0.10$&$-0.52$ & $-0.61$&$-0.42$ & $-0.40$\\
$\mu_L (\Sigma^{0})$&$ -1.61 \pm 0.10$& $-1.49$ &$-1.51$&$-1.59$ &$-1.62$\\ \hline
$\chi^2 /DoF$&&$2.14$&$1.09$&$2.47$&$2.05$ \\ \hline
Fitted &&$\mu_{u}=2.27$ &$\mu_{u}=2.28$&$\mu_{u}=2.40$ &$\mu_{u}=2.63$\\
Param &&$\langle S_{z} \rangle=0.26$ &$\langle S_{z} \rangle=-0.03$&
$\langle S_{z} \rangle=0.46$ &$\langle S_{z} \rangle=-0.33$\\
&& &$\langle L_{z} \rangle=0.23$  & &$\langle L_{z} \rangle =1.09 $ \\ \hline
\end{tabular}

\section*{ \normalsize  \bf TABLE II. Fits of magnetic moments when quark interactions are Coulomb
potential. }

\begin{tabular}{c|c|c|c|c|c} \hline
&Experimental & \multicolumn{2}{|c|}{Hypothesis I} & \multicolumn{2}{|c}{Hypothesis II}\\  \cline{3-6}
&Data & fit 1 & fit 2 & fit 1 & fit 2 \\ \hline
$\mu_C (p)$& $2.79 \pm 0.10$ & 2.76 & 2.71 &2.80 & 2.80 \\
$\mu_C (n)$& $-1.91 \pm 0.10$ &$-1.80$ & $-1.86$ &$-1.80$ & $-1.80$\\
$\mu_C (\Sigma^{+})$&$ 2.46 \pm 0.10 $&2.62 & 2.57 &2.62 & 2.61\\
$\mu_C (\Sigma^{-})$&$ -1.16 \pm 0.10$ &$-1.17$ & $-1.19$ &$-1.09$ & $-1.09$\\
$\mu_C (\Xi^{-})$&$ -0.65 \pm 0.10$&$-0.56$ & $-0.58$& $-0.48$ & $-0.48$\\
$\mu_C (\Xi^{0})$&$-1.25 \pm 0.10$&$-1.32$ & $-1.38$& $-1.37$ & $-1.38$\\
$\mu_L (\Lambda^{0})$&$-0.61 \pm 0.10$&$-0.56$ &$-0.41$&$-0.45$ & $-0.45$\\
$\mu_L (\Sigma^{0})$&$ -1.61 \pm 0.10$& $-1.49$ & $-1.51$&$-1.49$ & $-1.48$\\ \hline
$\chi^2 /DoF$&&$1.14$&$1.56$&$2.09$&$2.13$ \\ \hline
Fitted &&$\mu_{u}=2.22$ & $\mu_{u}=2.24$&$\mu_{u}=2.01$ & $\mu_{u}=1.99$\\
Param &&$\langle S_{z} \rangle=0.17$ & $\langle S_{z} \rangle=0.31$&
$\langle S_{z}\rangle=-0.30$ & $\langle S_{z} \rangle=0.86$\\
&&& $\langle L_{z} \rangle=0.04$ && $\langle L_{z} \rangle =-0.37$ \\ \hline
\end{tabular}

\end{document}